\documentclass[aps,showpacs,pra,twocolumn]{revtex4}

\usepackage{amsmath}
\usepackage{graphicx}
\usepackage{epsfig}
\newcommand{\be}{\begin{equation}}
\newcommand{\ee}{\end{equation}}
\newcommand{\bma}{\begin{displaymath}}
\newcommand{\ema}{\end{displaymath}}

\begin{document}

\title{Vortex localization in rotating clouds of bosons and fermions}

\author{S.M. Reimann$^1$, M. Koskinen$^2$, Y. Yu$^1$ and 
M. Manninen$^2$}

\affiliation{\sl $^2$Mathematical Physics, LTH, Lund University,
SE-22100 Lund, Sweden}

\affiliation{\sl $^1$NanoScience Center, Department of Physics,
FIN-40014 University of Jyv\"askyl\"a, Finland}

\begin{abstract} 
Finite quantal systems at high angular momenta may exhibit vortex formation 
and localization. These phenomena occur independent of the statistics of the 
repulsively interacting particles, which may be of bosonic or fermionic nature.
We analyze the relation between vortex localization and formation of 
stable Wigner molecules at high angular momenta in the view of 
particle-hole duality.
Trial wave functions for the vortex states and the corresponding 
fermion-boson relations are discussed. 
\end{abstract}

\maketitle

\section{Introduction}

When small quantal systems are set rotating, vortices may form. 
These vortices are characterized by 
rotational flow of particle current around minima in the density 
distribution. In contrast to fluid mechanics, a vortex in a quantum 
system  can be  $n$-fold quantized, with the 
corresponding wavefunction having a ($n$-fold) zero at the vortex position 
and a phase changing by $n 2\pi $ on a path around this zero.
 
The existence of a triangular vortex lattice in superconductors 
was predicted by Abrikosov~\cite{abrikosov1957} already in the 50ies. 
In the mesoscopic regime, finite-size effects determine the 
symmetry of multivortex configurations. 
In small systems, a compromise must be found between the triangular 
lattice of the bulk, and the shape of the confinement. It was 
predicted~\cite{schweigert1998,baelus2001,baelus2002,baelus2004}
that on a small superconducting disk, instead of the triangular Abrikosov 
lattice, the vortices may form 
geometric patterns that resemble those of classical point charges in a
harmonic potential, i.e. small, finite-size Wigner crystals or 
so-called Wigner molecules~\cite{bolton1993,bedanov1994}.
This was experimentally confirmed very recently by 
Grigorieva {\it et al.}~\cite{grigorieva2006}. 

However, the appearance of vortices and their localization is not limited to 
superconductivity or superfluidity.
Another well-known example is a  Bose-Einstein condensate
(BEC) of (repulsive) alkali-metallic atoms (see, for
example, Dalfovo {\it et al.}~\cite{dalfovo1999} and
Legett~\cite{leggett2001}, or the book by Pethick and Smith~\cite{pethick} 
for reviews on BEC), which is set rotating. 
With increasing angular momentum, an ever larger number of vortices 
may penetrate the bosonic cloud of atoms~\cite{madison2000,chevy2000,abo}. 

In BEC, much work focused on the Thomas-Fermi regime of strong 
coupling~\cite{dalfovo1999,rokshar1997,garcia1999,feder1999a,feder1999b,svidzinsky2000}, 
where the kinetic energy dominates. In the dilute limit, however,
even though the interactions are weak they play the major 
role~\cite{wilkin1998,butts1999,mottelson1999,bertsch1999,kavoulakis2000,wilkin2000,linn1999}.
Within the Gross-Pitaevskii mean field approach, Butts and 
Rokshar~\cite{butts1999} and later 
Kavoulakis {\it et al.}~\cite{kavoulakis2000} 
found geometric vortex configurations between distinct
values of angular momentum~\cite{vorov2003,vorov2005}. 
In analogy to mesoscopic superconducting disks, 
these vortex configurations again show  selected symmetries corresponding to 
those of small Wigner crystallites~\cite{manninen2005}. 

Surprisingly, vortices may form independent of the statistics
of the quantum particles. 
A striking identity  between boson and fermion many-particle configurations 
leads to similar vortex states for bosons as well as for fermions, implying 
that vortex formation indeed is a universal phenomenon in a rotating 
quantum system~\cite{toreblad2004}. 

At extreme angular momenta, far beyond the point 
where the vortex lattice melts~\cite{schweikhard2004}, 
bosonic~\cite{manninen2001,romanovsky2004,reimann2006} as well as 
the fermionic {\it particles} 
crystallize~\cite{jeon2004,jeon2004b,reimann2006}, 
approaching the classical limit~\cite{nikkarila2006}. 

The purpose of this article is to show 
how {\it vortex formation and localization}, as well as 
the {\it crystallization of particles} at the limit of extreme rotation 
for bosons {\it and } fermions are connected. 
Our analysis is largely based on the fact that 
vortex formation can most easily be analyzed in terms of particle-hole
duality, which was found to hold for the bosonic as well as the fermionic 
case~\cite{manninen2005}.

This paper is organized as follows: In section II, 
after describing the model and methods,
we briefly discuss the particle-hole dualism for the 
fermion vortex states.
Results of the direct numerical diagonalization of the 
many-body Hamiltonian follow in III, 
where we discuss the localization for bosons and fermions at high 
angular momenta in terms of the regular oscillations in the yrast line. 
We apply similar arguments to explain the localization 
of vortices, both in the bosonic and fermionic limit.
For fermions, this fact is further illustrated making use of 
particle-hole duality.  
Trial wave functions for the vortex states and the corresponding 
fermion-boson relations are discussed in section IV. 

\section{The Model}

\subsection{The Hamiltonian}
\label{sec_hamilton}

Let us now consider interacting particles confined by
a two-dimensional harmonic trap. 
Ignoring the spin degree of freedom, these particles can either 
be spinless bosons (as for example, bosonic atoms in the 
same hyperfine state),
or polarized spin-1/2 fermions (say, electrons).
The many-particle Hamiltonian is simply
\be
H=-\frac{\hbar^2}{2\mu}\sum_{i}^N \Delta _i 
+\sum_i^N \frac{1}{2}\mu\omega ^2 r_i^2
+\sum_{i<j}^N v(\vert {\bf r}_i-{\bf r}_j\vert)
\label{hamiltonian}
\ee
where $N$ is the number of particles with mass $\mu$, $\omega $ the 
oscillation frequency of the confining potential, and
$v(r)$ a {\it repulsive} two-body interaction. 
For a (dilute) gas of spinless, bosonic atoms,  often a contact 
interaction of the form  $v(r) = {1\over 2} U_0\delta (r)$
is used, where $U_0 $ then depends on the scattering 
length for the atom-atom collisions~\cite{kavoulakis2000}.
For spinless fermions, however, this interaction only recovers the
non-interacting case due to the Pauli principle.
To ease the direct comparison between the boson and fermion 
spectra, in both cases we thus consider the usual long-range Coulomb form,  
$v(r)={e^2}/{4\pi\epsilon_o r}$.
For angular momenta well below the fractional quantum Hall 
regime~\cite{wilkin1998,cooper1999,wilkin2000},
the boson spectra, calculated with either the 
short-range or long-range Coulomb interaction, in fact
show a remarkable similarity, 
as earlier demonstrated by Toreblad {\it et al.}~\cite{toreblad2004}.  
We refer to the work by Hussein {\it et al.}~\cite{hussein2002a,
hussein2002b, hussein2002c} and Vorov {\it et al.}~\cite{vorov2003}
for a more general discussion of the universality of repulsive 
interactions in the bosonic case.

In what follows, we are mainly interested in the general structure 
of the eigenstates of Eq.~\ref{hamiltonian} at high angular momentum $M$. 
We follow the nuclear 
physics tradition to study the so-called {\it yrast line}, i.e. the 
lowest-energy states as a function of $M$, and the corresponding 
low-lying excitations (the so-called 
{\it yrast spectrum}~\cite{grover1967,bohr1969}).
Similarly, magnetic fields are not included explicitly: 
neglecting the Zeeman term, the only effect a magnetic field has, 
is to set the system rotating.

\subsection{Restriction to the Lowest Landau Level}
\label{sec_LLL}

\begin{figure}[h]
\includegraphics[width=0.8\columnwidth]{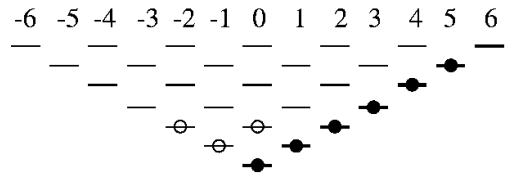}
\caption{Single-particle energy levels of a two-dimensional harmonic 
oscillator. The black bullets show the Maximum Density Droplet (MMD)
configuration for  six particles, and the open circles the $M=0$ 
ground state for noninteracting, 
polarized fermions. The lowest-energy levels for each $m$ form
the Lowest Landau Level (LLL), as indicated by the bold lines.
}
\label{spstates}
\end{figure}

In the absence of interactions, the single-particle energies of the 
two-dimensional harmonic oscillator are 
$\epsilon = \hbar \omega (2n+|m| +1),$
where $n$ is the radial quantum number, and $m$ the single-particle angular 
momentum.
The energy level structure is schematically sketched in Figure~\ref{spstates}. 
At very large total angular momentum of the non-interacting 
many-particle system, the lowest-energy state is characterized by 
quantum numbers $n=0$, 
and $m$ being zero or having the same sign as the total angular momentum, $M$.
Measured relative to the non-interacting ground state, the total energy of the
lowest state at a given $M$ thus equals $\hbar \omega  (M+1)$. 
This single-particle basis is identical to the so-called Lowest 
Landau Level (LLL) at strong magnetic fields. 
In this subspace, a configuration can be denoted by the Fock state
$\vert n_0 n_1 n_2 \cdots n_{s}\rangle$, where $n_i$ is the
occupation number for the single-particle state with angular momentum
$i$, and $s$ the largest single-particle angular momentum included
in the basis. For spinless fermions, $n_i$ is 0 or 1 
and for bosons it is an
integer. As the angular momentum $M$ is a good quantum number, 
we have the restriction $\sum_i i n_i=M$. 

For {fermions}, 
the smallest angular momentum that can be built in the LLL 
is that of the so called Maximum Density 
Droplet~\cite{macdonald1993}, $M_{MDD}=N(N-1)/2$,
where due to the Pauli principle the lowest-energy states with 
successive single-particle angular momenta are occupied, 
generating a compact configuration
\begin{equation}
\mid \underbrace {111 \dots 111}_N 000 \dots \rangle 
\label{MDD}
\end{equation}
(see Fig.~\ref{spstates}). 

\smallskip

To see the similarity with quantum Hall systems~\cite{reimann2002}, one may
define a {\it filling factor} $\nu $ by the ratio between $M_{MDD}$ and 
the actual angular momentum $M$,
\begin{equation}
\nu=N(N-1)/(2M)~.
\end{equation}
Clearly, the MDD corresponds to the Laughlin quantum Hall state at
filling factor one in the bulk~\cite{laughlin1983}. 
For {bosons}, the state with 
smallest possible angular momentum within the LLL is the 
{\it non-rotating}  
ground state, where all particles occupy the single quantum 
state with $n=0$ and $m=0$, i.e., $\mid N 0 0 0 \dots \rangle $. 

\subsection{Numerical diagonalization of the Hamiltonian}
\label{sec_numdiag}

Including now the interactions between the particles, 
we solve the full many-particle Hamiltonian  exactly by numerical
diagonalization, in order to obtain the ground state and the low-lying 
excited states at a given angular momentum $M$. In this so-called 
configuration interaction (CI) method,  the many-particle state is 
described as a linear combination of Slater determinants, 
$| \Psi \rangle = \sum C_{\alpha } |{\cal L }_{\alpha }\rangle $, which for given 
statistics are defined by the combinatorics of the single-particle basis.
Here, it is natural to use the basis
of the Harmonic oscillator (i.e. that of the external trap).
If the strength of the interparticle interaction is much smaller
than the single-particle excitation energy $\hbar\omega $, 
at large $M$  the most
important configurations indeed consist of single-particle states of the
LLL. In the limit $e^2/4\pi\epsilon_0 r_s\hbar\omega_0\rightarrow 0$ 
(or $v_0/\hbar\omega_0\rightarrow 0$),  where $r_s$ is the 
electron density parameter,  $n_0=1/\pi r_s^2$,  configurations 
with particles at higher Landau levels have diminished 
weight $C_{\alpha }$, and the LLL approaches the exact result. 
(For a discussion that includes Landau level mixing in
the small-$N$ limit, see the recent work by  
Stopa {\it et al.}~\cite{stopa2006}, as well 
as G\"u{\c c}l\"u {\it et al.}~\cite{guclu2005,guclu2005b}, 
considering quantum dots at high magnetic fields. A similar study was carried 
out for trapped bosons with contact interactions by Morris and 
Feder~\cite{morris2006}).

Increasing the angular momentum beyond the MDD in the fermionic case,
$M> M_{MDD}$, or beyond the ``condensate'',  $M>0$, in the bosonic case, 
there are many ways to distribute $M$ quanta of angular momentum on 
the given number of particles, $N$. 
For large values of $N$ and $M$, we thus have to restrict the basis 
by setting a smaller value $s$ for the largest single-particle angular
momentum. However, the 
most dominant configurations appear rather compact, i.e. 
most of the occupied single-particle states are close in angular momentum
$m$. Thus, accurate results can still be obtained in a
restricted Hilbert space. 

In the LLL the single-particle basis states
in polar coordinates $(r,\phi)$ are
\be
\psi_m(r,\phi)=A_m r^me^{-r^2/4}e^{im\phi},
\label{spstate}
\ee
where $A_m$ is a normalization factor.

The many-particle Hamiltonian, Eq.~(\ref{hamiltonian}),  can be written,
apart from a constant, as 
\be
H=\sum_i m_i\hbar\omega c_i^+c_i +\sum_{i,j,k,l}
V_{ijkl} c_i^+c_j^+c_kc_l,
\label{sqH}
\ee
where
\be
V_{ijkl}=\int\int d{\bf r}d{\bf r'}
v(\vert{\bf r}-{\bf r'}\vert)\psi_i^*({\bf r})\psi_j^*({\bf r'})
\psi_k({\bf r})\psi_l({\bf r'}),
\label{vijkl}
\ee
and $v$ is the interparticle interaction (see above).
Since the total angular momentum is fixed, $M=\sum m_i$,
the diagonal part gives an energy $\hbar\omega M$ for
all configurations. In effect, we thus only 
diagonalize the interaction part of the Hamiltonian.
This simplification is caused by the restriction of the basis
to the lowest Landau level and holds for 
bosons as well as for fermions. It is natural to present the
energies of the interaction part only (which trivially scales 
with its strength), as done for most of the results shown below.

The interaction matrix elements are computed numerically. 
In the case of Coulomb interaction
we used the technique suggested by Stone {\it et al.}~\cite{stone}. 
For numerical diagonalization of the Hamiltonian matrix, we applied the 
Lanczos method~\cite{arpack}.

\subsection{Particle-hole duality}
\label{sec_holes}

A Fock state can be described either by  particles,  or
equivalently,  by holes. For spinless fermions, 
in practice one simply can replace the zeros with ones, and vice versa the
ones with zeros.
For example, the configuration $\vert 1110011111\rangle$ is then changed to
$\vert 0001100000\rangle $.
One correspondingly  defines the creation (annihilation)
operator of a hole as $d^+=c$ ($d=c^+$) and writes
the Hamiltonian Eq.~(\ref{sqH}) in terms of these:
\begin{eqnarray}
\label{sqHh}
H&=&\sum_i m_i\hbar\omega_0(1-d_i^+d_i)
+2\sum_{i,j}\left(V_{ijij}-V_{ijji}\right)d_i^+d_i \nonumber \\ 
&+&\sum_{i,j,k,l}V_{ijkl}d_l^+d_k^+d_jd_i+{\rm constant}.
\end{eqnarray}
Naturally, the solution of this Hamiltonian leads to an 
equivalent result, and requires the same numerical effort. 

The interactions between the holes here are the same
as the interactions between the particles. 
However, it is important to note that the second term
is nonzero. This means that the holes effectively do not 
move in a harmonic confinement. This has two
important consequences: the holes ($i$) do not 
have a pure center of mass excitation, and 
($ii$) not necessarily the same localization geometry as the particles.

This particle-hole dualism, as earlier described in Ref.~\cite{manninen2005},  
allows us to identify the holes as vortices in the Fermi sea and
to evaluate the correlation functions between these
vortices. Furthermore, in the case of a few holes in the Fermi sea, the
energy spectrum is dominated by these holes
and can be understood by diagonalizing only
the interaction part of the hole Hamiltonian
(i.e. the third term in Eq.~(\ref{sqHh})).

\section{Results}

\subsection{Localization of particles at extreme angular momenta}
\label{sec_res:loc}

Let us first analyze the many-body quantum spectra 
for small particle numbers. (This will later turn useful when 
studying the spectral properties of many-particle
systems in terms of localized vortices).

\begin{figure}[h]
\includegraphics[width=1.1\columnwidth]{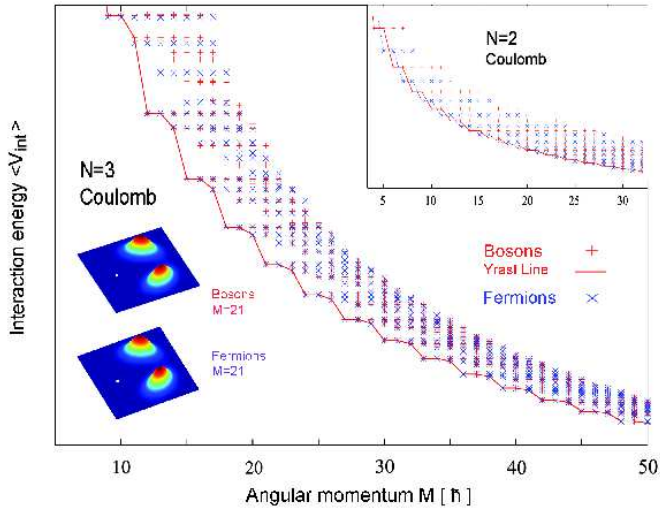}
\caption{Interaction energy of a three-particle system, $N=3$, for 
bosons ({\it red crosses,}~+) and fermions ({\it blue crosses},~$\times $), 
as a function of the angular momentum, for angular momenta 
$M>3M_{MDD}$~. (The inset to the right shows the 
same for $N=2$). The red (blue) line is the yrast line.
The insets to the left show the boson and fermion correlation functions, 
here taken at the cusp states with $M=21$ as an example.  
The white dot indicates the reference point.
The particles appear localized at the corners of a triangle, with one particle 
placed at the reference point.} 
\label{spect3}
\end{figure}

Figure~\ref{spect3} compares the spectra for three bosons
({\it red crosses},~$+$) and three fermions ({\it blue crosses},~$\times $)
at extreme angular momenta,~$M>3M_{MDD}$.
(Only the nontrivial  interaction part of the total energy is shown, here 
for the ten lowest states at fixed $M$).
Naturally, the interaction energy goes down with increasing angular momentum 
for repulsive interactions. Note that for each
new energy at a given value $M^\prime $, the same energy is reproduced at 
all angular momenta $M> M^\prime $. These are simply the 
center of mass excitations characteristic for harmonic 
confinement~\cite{trugman1985}.
With increasing $M$, the boson and fermion spectra become remarkably
similar. This does not only hold for the yrast line, but also 
for the low-lying excitations.

\begin{figure}[h]
\includegraphics[width=1.1\columnwidth]{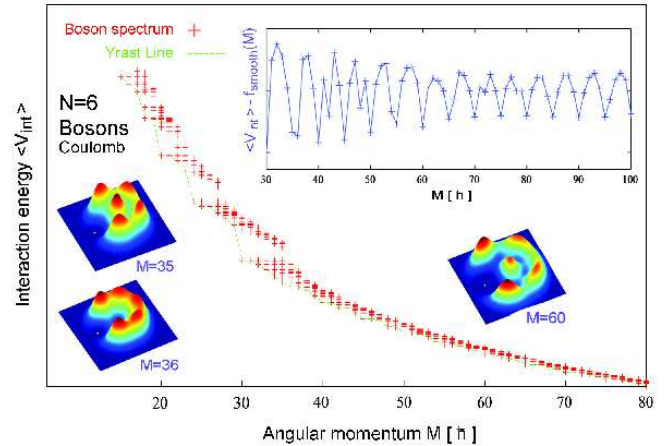}
\caption{Interaction energy of $N=6$ bosons 
as a function of the angular momentum. 
The inset shows the yrast line with a smooth function of angular 
momentum (3rd order polynomial) subtracted from the 
energies, in order to make the oscillations more visible. 
The large-$M$ limit is dominated by a regular oscillation with $\Delta M=5$. 
The pair correlation functions to the left 
clearly demonstrate localization in Wigner molecule geometries
at high angular momenta. 
While at smaller $M$-values, the 
$(1,5)$ and the $(0,6)$ configurations compete, at extreme angular momenta 
fivefold symmetry dominates.}
\label{spect6}
\end{figure}
The yrast line (drawn in Fig.~\ref{spect3} as a red line to
guide the eye) shows clear oscillations with a period of $\Delta M= 3$.
The inset shows the corresponding spectrum for $N=2$, with a similar 
oscillation in the yrast states, alternating between even and odd angular
momenta, respectively. 
Further increasing the number of particles, we show in Fig.~\ref{spect6}
the yrast spectrum for $N=6$ bosons.
Again, the boson and fermion spectra (not plotted here) 
are very similar in the large-$M$ limit. 

Similar to the results shown for $N=3$ in Fig.~\ref{spect3} above, 
the boson spectrum in Fig.~\ref{spect6} shows pronounced cusps with a 
characteristic gap to the low-lying excitations. At smaller $M$, 
oscillation periods with $\Delta M= 5$ and $\Delta M=6$ compete,
with cusps at $M=18, 20, 24, 25, 30, 36, 40...$. 
Beyond $M=50$ the cusps appear regularly with $\Delta M=6$. 
This is very clearly seen in the inset to Fig.~\ref{spect6}, 
where a third order polynomial in $M$ 
was subtracted from the energies, changing the
slope of the yrast line in order to make its oscillations more visible.

These cusp states and their periodicities in $M$ are a phenomenon well
known to occur in quantum dots 
at very strong magnetic fields, which have been investigated 
extensively in the literature~\cite{reimann2002}.
Here, the regular oscillations of ground state energy as a function of 
magnetic field were found to originate from rigid rotations of the 
classical electron configurations~\cite{maksym1990,maksym1996}. 
While for particle numbers up to $N=5$, the Wigner crystallites form 
simple polygons (from triangle to 
pentagon), the case $N=6$ is more complicated due to 
the interplay between two stable classical geometries, a pentagon with 
one particle at the center, $(1,5)$, and a 
hexagon, $(0,6)$~\cite{manninen2001}.
 
These simple geometries in fact easily explain 
the periodic oscillations in the above many-body spectra. 
The rigid rotation of the 'molecule' of localized particles is only possible
at angular momenta allowed by the underlying symmetry group~\cite{reimann2006}.
For example, if the electrons localize for example in an equilateral triangle, 
the three-fold symmetry leads to period of three.  The five-fold 
symmetry of the $(1,5)$ ground state of the six-electron molecule yields a 
period of five~\cite{koskinen2001,manninen2001,reimann2002,viefers2004}.
Correspondingly, the period $\Delta M=6$ is explained by the 
hexagonal structure, $(0,6)$.

A related periodicity of the yrast energy as a function of the 
angular momentum is obtained also for other particle numbers, and holds 
even if higher Landau levels are included in the 
basis set~\cite{maksym1990,maksym1996,wojs1997,manninen2001}.
(Similar agruments were successfully applied earlier also 
to analyze the rotational spectrum  of 
quasi-one-dimensional quantum rings, see~\cite{koskinen2001,viefers2004}).

The particle localization in the internal coordinates of the quantum 
system  can be demonstrated further by using a 
rotating frame~\cite{maksym1996}, or 
by studying the pair correlations defined as 
\be
g({\bf r},{\bf r'})=
\langle \Psi\vert \hat n({\bf r})\hat n({\bf r'})\vert \Psi\rangle ~,
\label{g}
\ee
where $\hat n$ is the density operator. The pair correlation
function $g$ describes the probability to find a particle at
${\bf r'}$ when another particle is in the reference point ${\bf r}$.
In a finite system, the pair correlation function is 
a function of two variables, and often  
also called ``conditional probability''. 
In Fig.~\ref{spect3} ({\it left}) we show the pair correlations 
for the three-particle system, both in the bosonic and fermionic case, 
at $M=21$. In both cases, the particles appear localized in  a triangular 
geometry, with two particles showing clear maxima opposite to the 
reference point (indicated by the white dot).  
A very similar result is obtained for the six-particle system, 
where at smaller $M$-values, the 
$(1,5)$ and the $(0,6)$ configurations compete. At extreme angular momenta, 
however, as here shown for $M=60$,  the fivefold symmetry $(1,5)$ dominates.
These pair correlations clearly demonstrate that at high angular momenta,  
bosons as well as fermions 
localize in Wigner molecule geometries, in excellent agreement 
with the earlier results by Manninen {\it et al.}~\cite{manninen2001},  
and their subsequent confirmation by 
Romanovsky {\it et al.}~\cite{romanovsky2004}.  

\subsection{Localization of Vortices}
\label{sec_20electrons}

For a harmonically trapped cloud of bosons that is set moderately rotating, 
geometric vortex configurations are known to appear between distinct
values of angular momentum. A single vortex at the center of the system
is formed when $M=N$, two vortices 
at about $M=1.7N$, three vortices when $M=2.1N$ and
$N$ vortices before $M=3N$. 
It was shown earlier that these mean-field results within the 
Gross-Pitaevskii scheme~\cite{butts1999,kavoulakis2000} 
emerge as the correct leading order approximation 
to exact calculations in the same subspace~\cite{jackson2001}.  

\begin{figure}[h]
\includegraphics[width=0.9\columnwidth]{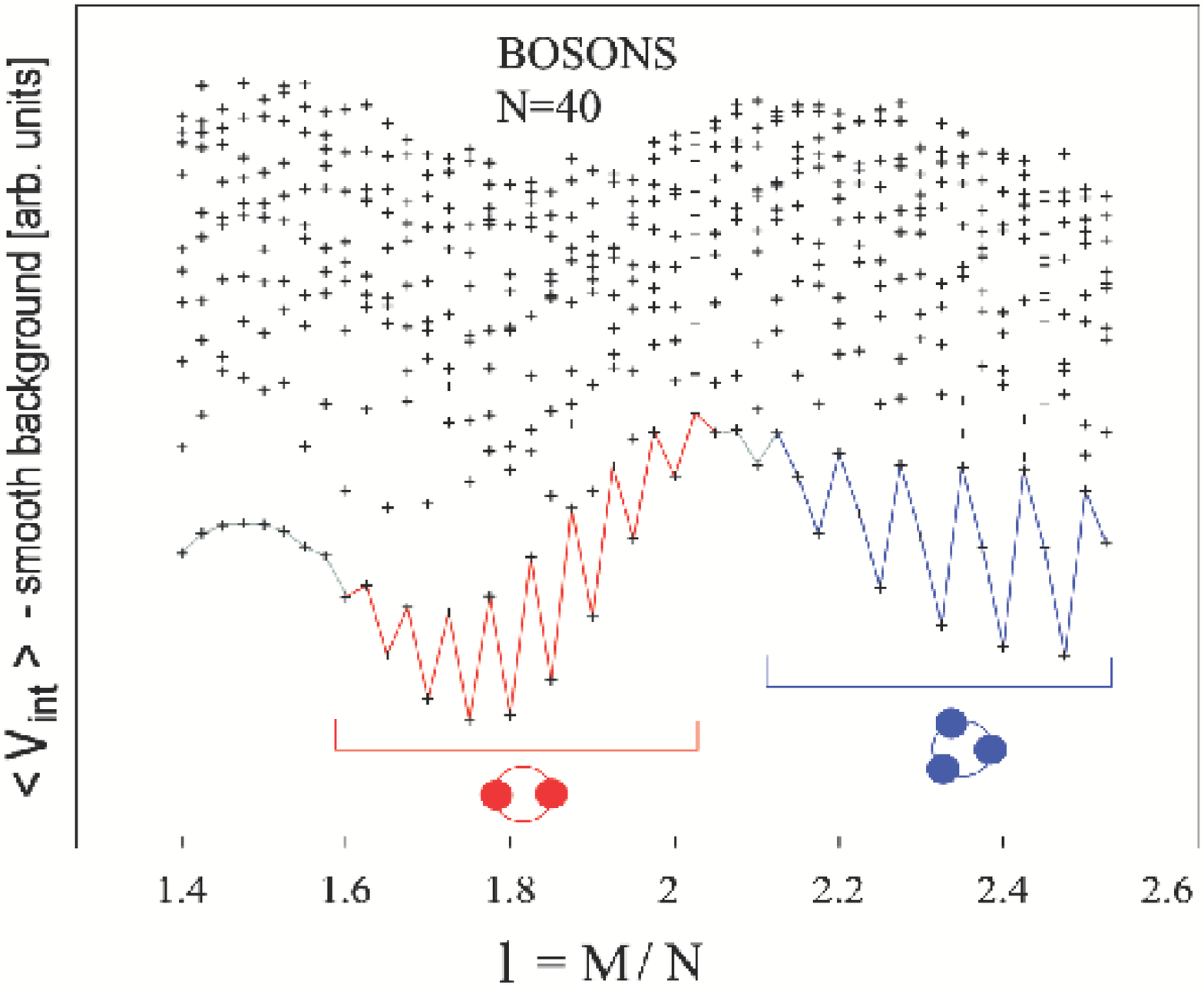}
\includegraphics[width=0.9\columnwidth ]{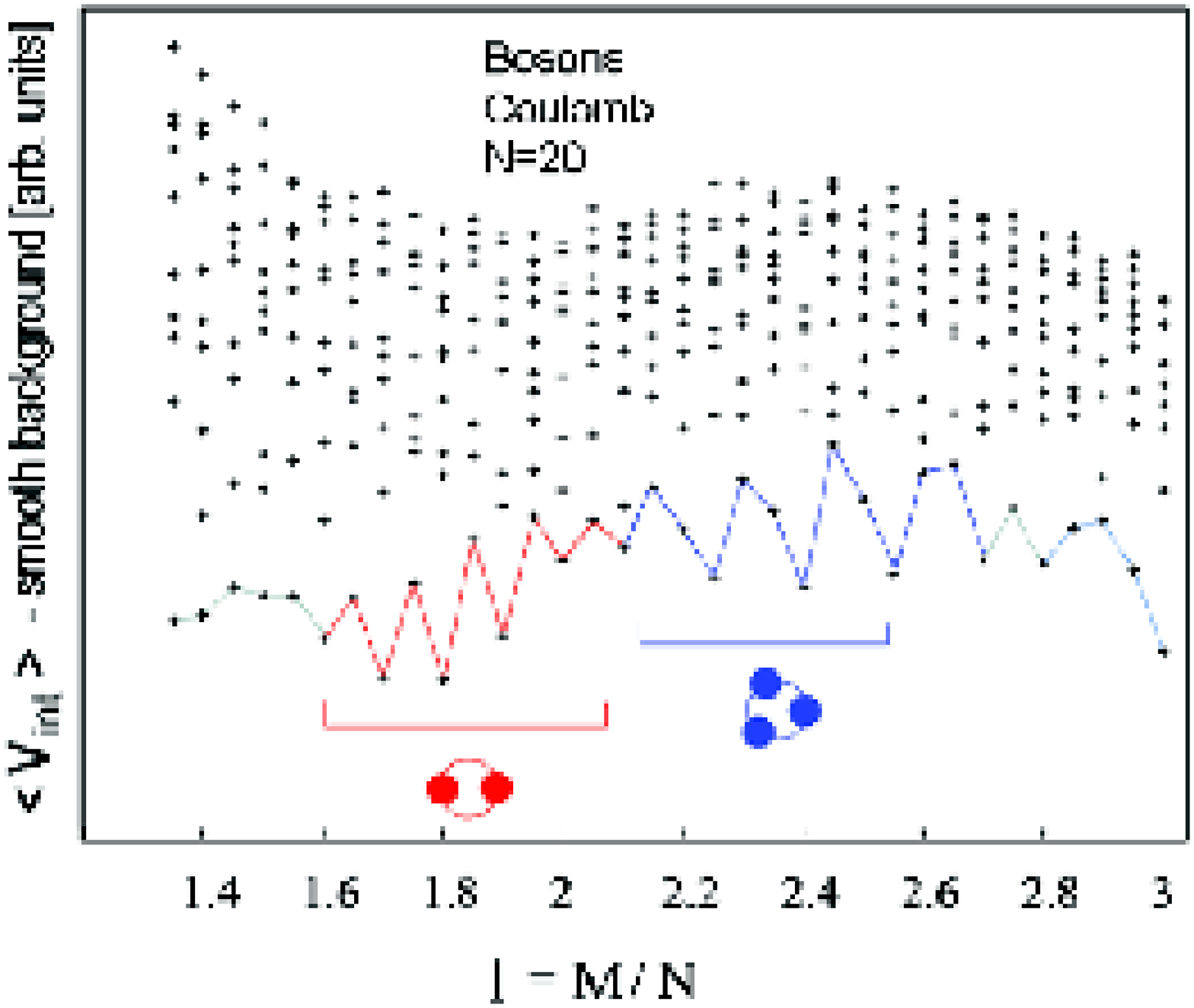}
\caption{
Many-particle energy levels for 20 ({\it lower panel}) and 
40 bosons ({\it upper panel}) as a function of angular momentum
per particle, $l=L/M$. In both cases, a second-order polynomial in $M$ 
was subtracted from the energies to make the yrast line
nearly horizontal. 
}
\label{n20n40b}
\end{figure}

The many-body spectra for 20 and 40 bosons interacting by the Coulomb 
force, are compared in Figure~\ref{n20n40b}, displaying  the yrast line and 
up to ten low-lying excitations. The horizontal axis is
now the angular momentum per particle, $l=M/N$. 
As previously, a second-order polynomial in $M$ was subtracted 
to emphasize the salient oscillations of the yrast line. 
For these large particle numbers, the Hilbert space was truncated 
such that only single-particle states with $m\le 10 $ were included, 
still giving well-converged results in the bosonic case. 

Both spectra show pronounced 
oscillations with consecutive periods of $\Delta M= 2, 3$ and 4, 
expanding through the above mentioned intervals of subsequent vortex 
entry. While a single vortex is formed at $l=1$, 
oscillations with $\Delta M=2$ coincide with the two-vortex solution appearing 
between $1.7< l < 2.1 $, three vortices beyond $l=2.1$, etc. 
These results are in excellent agreement with the results of
the Gross-Pitaevskii approximation~\cite{kavoulakis2000}.

The regular oscillations in fact appear very similar to the 
cusp states at extreme angular momentum, as discussed 
for small particle numbers 
in Figs.~\ref{spect3}  and ~\ref{spect6} above. However, 
in contrast to localization in Wigner molecules, these oscillation 
are now caused by two, three, or four localized {\it vortices}, respectively.
(The 2- and 3-vortex geometries are schematically 
sketched in Figs.~\ref{n20n40b} and~\ref{spect20} below). 

We notice that the spectra look qualitatively
very similar to that for 20 fermions shown in 
Fig.~\ref{spect20}. There is, however, a marked
quantitative difference in the angular momentum where 
the vortices appear. In the case of bosons,
the value of angular momentum per particle $M/N$ where the second,
third, etc. vortex enters the cloud,  seems to be independent 
of the number of bosons in the system. (Naturally the 
number of particles has to be much larger than the number
of vortices). 
If the same
systematics would hold for fermions, we should expect 
the same small number of vortices to appear when the angular momentum is
increased beyond the MDD.  This, however, does not hold for $N\ge 12$.
The vortex systematics then changes~\cite{yang2002,saarikoski2005,toreblad2006}
as compared to the bosonic case. 
The increase of angular momentum per particle, $(M-M_{MDD})/N$,
for the appearance of the second vortex depends on $N$, 
as illustrated in Table~\ref{vformation}.
\begin{table}[h]
\caption{Angular momenta where the vortices appear for 
boson and fermion systems of different sizes. $N$ is the
number of particles, $M$ the angular momentum and 
${\bar M~}=M-M_{MDD}$}
\label{vformation}
\begin{tabular}{r|rrr|rr}
\hline
       & $[M/N]$ &for bosons & & $[{\bar M~}/N]$ & for fermions \\
vortex &  $N=8$      & $N=20$ &$N=40$   & $N=8$           &  $N=20$ \\
\hline
2nd    & 1.7          & 1.6   & 1.6       &  1.7  & 0.9 \\
3rd    & 2.2          & 2.1   & 2.1       &  2.2  & 1.4 \\
4th    &              & 2.8   & 2.8       &       & 2.2 \\
\hline
\end{tabular}
\end{table}

\begin{figure}[h]
\includegraphics[width=0.9\columnwidth ]{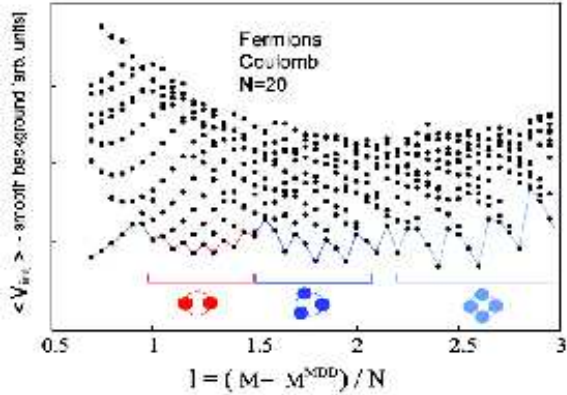}
\caption{Many-particle energy levels for 20 fermions
as a function of angular momentum per particle relative to the MDD, 
$l=(M-M_{MDD})/N$. As in Fig.~\ref{n20n40b} above, a second-order polyonomial 
was subtracted from the yrast line.
}
\label{spect20}
\end{figure}

The origin of the periodic oscillations can be understood 
by studying the corresponding many-particle problem
of holes, as briefly discussed in section~\ref{sec_holes} above.
 
For this purpose, let us now study in more detail the fermion case. 
The excitations from the MDD to higher angular
momenta create holes in the filled Fermi sea.
These holes are associated with the existence of  vortices
in the system~\cite{manninen2005}. 
In the single-vortex region, the most important
configuration has a single hole 
at angular momentum $M_h=M_{\rm MDD}+N-M$.
When $M$ is increased from the MDD the 
single vortex enters the system from the surface 
and reaches the center at angular momentum $M=M_{\rm MDD}+N$.
Then, $M_h=0$, i.e. the Fock state with 
the largest amplitude has the structure 
$\mid 0 111\dots 1111111000\dots \rangle $. 
This state appears in the excitation spectrum 
and terminates the band~\cite{toreblad2006}:
Within the lowest Landau 
level, no further quanta of angular momentum can be added, 
unless a second hole penetrates the electron droplet. 
Note that the single hole has a simple single-particle wave function, 
as in Eq.~(\ref{spstate}), i.e. the hole (or vortex) is
either localized at the center, or delocalized on a ring.

The single vortex does not reach the origin before it
is energetically more favorable to create two vortices
that are closer to the surface.  
Considering the state for holes, the angular momentum is written as
\begin{equation}
M_h=\frac{(N+h)(N+h-1)}{2}-M,
\label{Mh}
\end{equation}
where $h$ is the number of holes and $M$ the angular momentum of the 
fermion system.
For example, the 20-particle state with angular momentum
$M=M_{MDD}+24=214$ corresponds to a two-hole state with
$M_h=17$. The two-hole state at such high angular
momentum is strongly correlated, with holes localizing in
a narrow ring,  as shown in Fig.~\ref{twoel} below.

\medskip

Similar to the particle-particle ($pp$) correlations, we can define a 
hole-hole ($hh$) correlation function
\be
g_{hh}({\bf r},{\bf r'})=
\langle \Psi\vert \hat h({\bf r})\hat h({\bf r'})\vert \Psi\rangle,
\label{gh}
\ee
where $\hat h({\bf r})=\sum_{i,j} \psi_i^*({\bf r})\psi_j({\bf r})d_i^+d_j$
is the hole density operator. 

The pair correlation between the holes
clearly shows that in the internal frame,
the holes indeed are localized.
Fig.~\ref{twoel} shows the that the density profile of the two holes
forms a narrow ring. At the same radius, the $hh$-correlation
shows a very well localized maximum, corresponding to a hole (vortex)
localized at the opposite side of the ``reference hole''.

\begin{figure}[h]
\includegraphics[width=0.9\columnwidth]{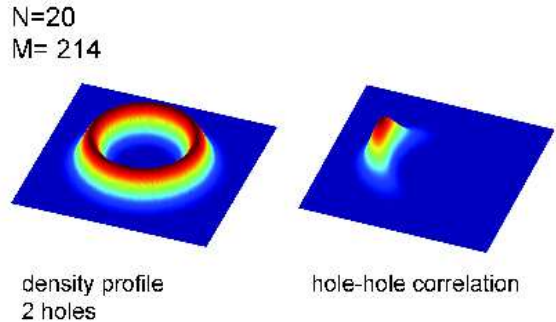}
\caption{Density profile of two holes at angular
momentum $M_h=17$ (a) and the corresponding hole-hole correlations (b). 
The hole density and pair
correlation was determined from the calculation for 20 electrons
with angular momentum $M=214$ and the minimum single-particle basis
allowing that angular momentum.
}
\label{twoel}
\end{figure}

Considering the two-hole system as a rigid ``molecule'' of two vortices,
the oscillations in the energy spectrum consequently are 
understood as resulting from the rotational states: Like 
for a two-atomic molecule, only every second angular
momentum is allowed in the rotational spectrum~\cite{tinkham1964}.
At angular momenta in between, the rotational state must be
occupied by a vibrational excitation. Thus, the state has 
higher energy. Similarly, this holds for the periodic oscillations 
of the two-particle yrast line for bosons as well as fermions,  
as displayed in the inset to Fig.~\ref{spect3} above. 

Increasing the angular momentum, the whole system expands, while the 
vortices move closer to the center. Beyond a certain angular momentum,
however, the repulsion 
between the vortices becomes so strong that it is energetically favorable to 
form a larger ring with three vortices instead. In this region,  
the spectrum (Fig.~\ref{spect20}) then 
shows an oscillation  period with $\Delta M=3$,
caused by the rigid rotation of the triangle of {\it three}
vortices. Further increasing the angular momentum adds additional vortices, 
with the oscillation period increased by one each time a new vortex enters
the system.

\begin{figure}[h]
\includegraphics[width=0.95\columnwidth]{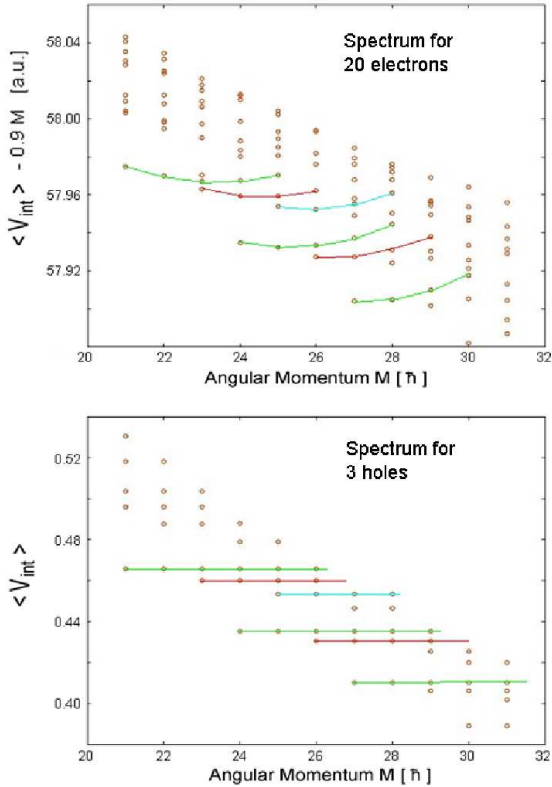}
\caption{(a) Many-particle spectrum for three holes, as
determined from the 20-fermion spectrum (see text). A linear function $ 0.9M$
was subtracted. The hole spectrum compares well with that of 
three {\it particles}, shown in (b) for Coulomb interactions. 
The horizontal lines in (b) connect states which are center-of-mass 
excitations. To guide the eye, corresponding lines are plotted in (a).
}
\label{spec20and3}
\end{figure}

Let us now study in more detail the three-vortex region, and   
compare the spectrum calculated for only three fermions, as shown in 
Fig.~\ref{spect3}, with that 
for 20 fermions, as displayed in the lower panel of Fig.~\ref{spec20and3}.
(For a quantitative comparison of the spectra, we
first subtracted from the 20-fermion spectrum a linear function
$0.9M$ and then plot it as a function of the corresponding 
hole angular momentum as defined in Eq.~(\ref{Mh})).

The spectra are strikingly similar,
indicating that indeed the spectrum of 20 particles (in this
region) is determined by the three localized holes.
The lines in Fig.~\ref{spec20and3} indicate the center of mass 
excitations for three particles. In the 20-particle spectrum,
these states are not pure center of mass
excitations of holes, since the true Hamiltonian for holes,
Eq.~(\ref{sqHh}), has additional terms with
anharmonic corrections to the confining potential.

\begin{figure}[h]
\includegraphics[angle=-90,width=0.95\columnwidth]{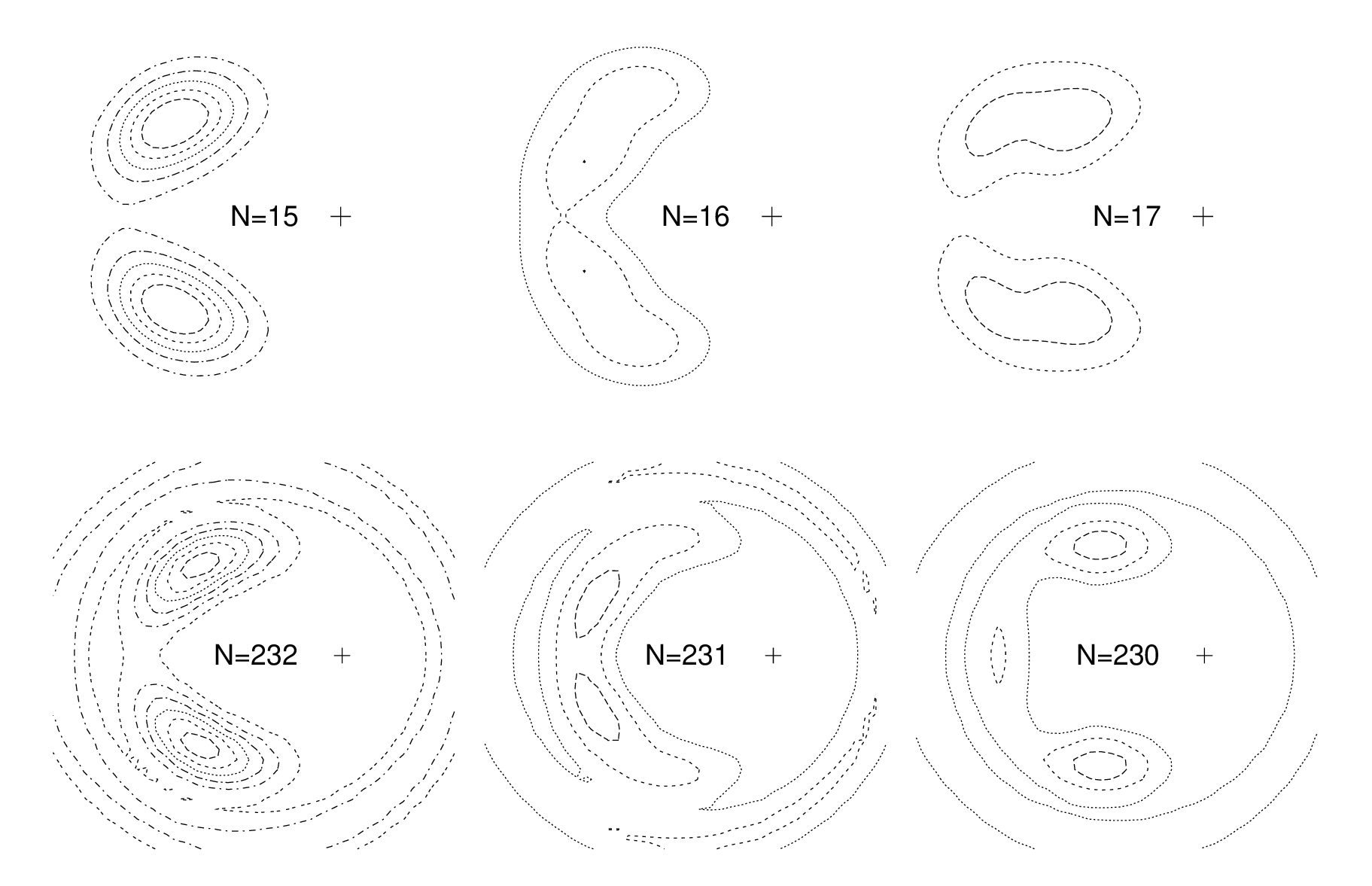}
\caption{The {\it lower panel} shows contours of the 
$hh$-correlations for three-vortex states of a 20-fermion 
system (Coulomb) at angular momenta 
$M=230$, $M=231$ and $M=232$. The {\it upper panel} shows the 
corresponding $pp$-correlations for {\it three} particles, 
at $M=15$, 16 and 17.  
}
\label{pair3}
\end{figure}

The observation that these quantum states show much similarity 
in their internal structure is further clarified by  comparing their 
pair correlations. Fig.~\ref{pair3}
shows the $pp$ correlations at three
points along the yrast line. When the yrast line has a kink, 
as here for example at $M_h=15$ or $M=239$,  the 
correlation function clearly shows localization
of holes in a triangular geometry. At $M_h=16$ the three-particle
system originated from the $M_h=15$ state  by 
a center of mass excitation, and at $M_h=17$ it has an
internal vibrational mode~\cite{reimann2006,nikkarila2006}.
The effect of these excitations is clearly seen in the correlation functions. 
The correspondence between $hh$ and $pp$ correlations for $N=3$ and $N=20$,
is not perfect, however, since in the 20-particle system 
the center of mass excitation (for holes) is mixed with the 
vibrational excitation.

\begin{figure}[h]
\includegraphics[width=0.7\columnwidth]{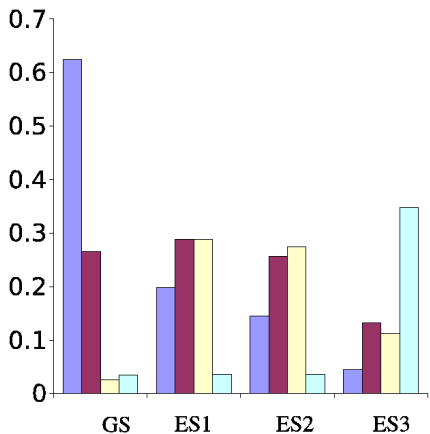}
\caption{Overlaps between 3-hole states from two
different calculations for $M_h=24$ ($M=22$). 
GS, ES1, ES2 and ES3 refer to the ground
state, and excited states of the three particle calculation.
The columns show the overlaps with the ground state (blue),
and the first three excited states (red, yellow and green) of
the hole state derived from the 20 electron-calculation.  
}
\label{overlaps}
\end{figure}

The three-particle and the three-hole states for $N=20$ can be 
compared directly by calculating their overlaps. In doing so, the 
electron calculation has to be restricted to the minimum 
Hilbert space having only three holes. As an example, we 
calculated the overlap matrix for the four lowest energy states
at angular momentum $M=24$ for the holes (the corresponding 20 
electron angular momentum is 229). The result is shown in 
Fig.~\ref{overlaps}. The overlap between the ground states of
the two calculations is large,  showing again that their internal
structure is similar. The first and second excited states 
seem to be mixed, while the third excited state is again quite
similar. This result is in agreement with the pair
correlation functions (Fig.~\ref{pair3}) and the spectra 
(Fig.~\ref{spec20and3}) which show that a pure center of mass
excitation of holes (first excited state here) does not exist in
the 20-particle spectrum. 

For small particle numbers, $N\stackrel {<}{\sim } 10$,  
the pair correlations cannot display the vortex structure properly. 
In this case, weak signals of vortices can be seen only if the 
reference point is clearly outside the density 
distribution~\cite{kavoulakis2002},
since then the exchange-correlation hole is distributed rather 
evenly over the remaining dot area. 
\begin{figure}[h]
\includegraphics[width=0.95\columnwidth]{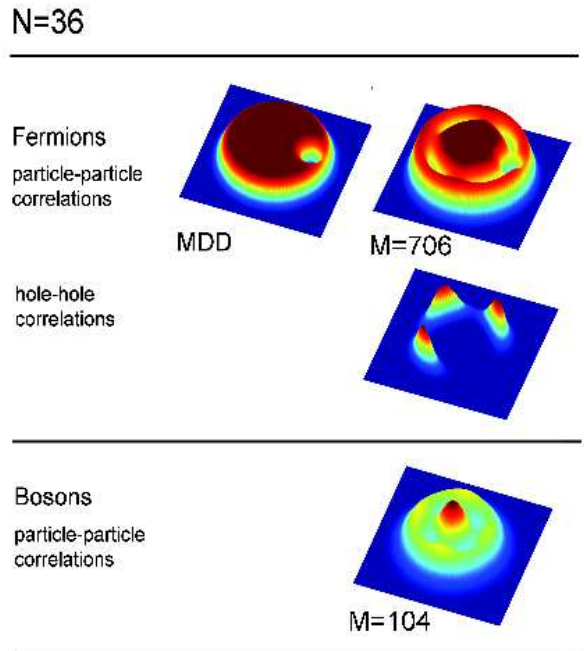}
\caption{Pair correlation functions calculated for 36 electrons.
The {\it upper panel} shows the  electron-electron correlations for the 
MMD ({\it left}), for particles at 
$M=706$ showing four vortices ({\it right}), and for holes at the same angular
momentum ({\it lower right}). 
The {\it lower panel} shows the corresponding  
correlation function for a  bosonic four-vortex state 
at angular momentum $M=104$. (Note the absence of the exchange hole in the
bosonic case.) 
}
\label{v4epc}
\end{figure}

The situation becomes different, when the particle number is 
significantly increased. Figure~\ref{v4epc} shows the electron-electron
pair correlation for 36 electrons. In the MDD, only the exchange hole 
at the reference point is clearly visible. 
In the case of $M=706$, in addition to the exchange hole we see 
four wide minima, which we interprete as four localized vortices. 
The reference point is chosen at
a radius where the density has a minimum, i.e.
at the expected radius of the vortex ring. 
The $hh$-correlation, also shown in Fig.~\ref{v4epc}, shows three 
pronounced maxima, which are consistent with the minima in the 
$pp$-correlation.
This comparison demonstrates, that the vortex localization 
is more clearly seen in the $hh$-correlations, than in the $ee$-correlations.
The lowest panel in Fig.~\ref{v4epc} shows the corresponding pair correlations 
in the bosonic case. 
Also in this case, we see four minima, corresponding to the four vortices, 
but naturally, not the exchange hole. 

\begin{figure}[h]
\includegraphics[width=0.9\columnwidth]{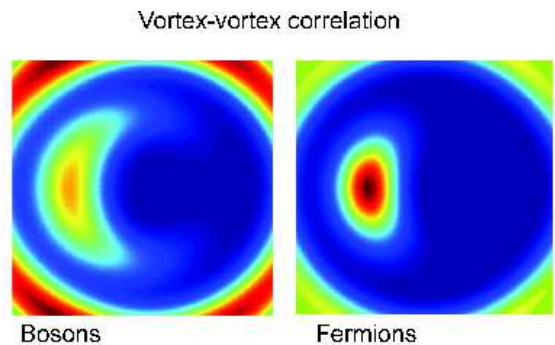}
\caption{Vortex-vortex correlation function
of the two-vortex state for 20 bosons with $M=34$, $M/N=1.7$,
left, and for 20 fermions with $M=M_{MDD}+34=224$.
The fermion state is the second excited state for that
angular momentum. 
For bosons the correlation function is 
determined by transforming the boson state
to a fermions state and plotting the hole-hole correlation. 
}
\label{bosfer2v}
\end{figure}

Regarding the analysis of the corresponding Fock states in the bosonic 
case, it is not as straightforward to relate the vortices
to holes in the similar fashion as for spinless fermions. 
Nevertheless, we can transform the boson wave function 
to a fermion wave function using the method explained 
in Ref.~\cite{toreblad2004}
and then plot the vortex-vortex correlation for the corresponding
fermion system. Figure~\ref{bosfer2v} shows a pair correlation 
function constructed in this way, with the 
localization of the second vortex opposite to the
reference point.

\subsection{Comparison between boson and fermion wave functions}
\label{sec_wfcompare}

\begin{figure}[h]
\includegraphics[width=0.9\columnwidth]{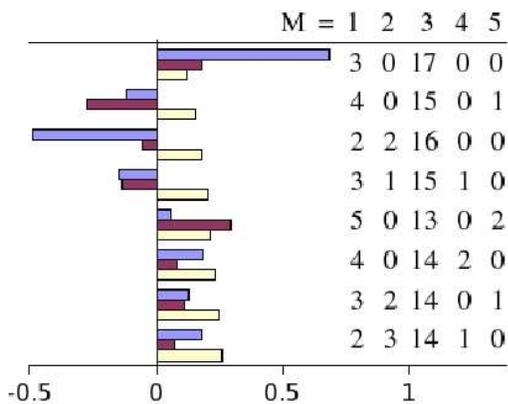}
\caption{Comparison of the coefficients of the most important
configurations 
of the two-vortex state of 20 bosons with
$M=34$ to those of 20 fermions with $M=224$. In the case of fermions
the state is the second excited state for the chosen angular momentum.
For fermions, the amplitudes are shown in blue, and for bosons in red.
The yellow columns show the cumulative overlap between the 
boson and fermion states. The configurations are shown at the right, 
using the notation for bosons.
}
\label{amps}
\end{figure}

Comparing  the energy spectra and the pair correlation functions, 
the vortex formation for bosons and 
fermions appears surprisingly similar. 
It is now interesting to see how far this similarity 
is reflected in the detailed structure of the many-particle states.  
For this purpose, we need to  
compare the fermion wave function with angular momentum
$M-M_{MDD}$ to the boson wave function with $M$. 
To this end, we should multiply the boson wave function with
the determinant $\prod (z_i-z_j)$. 
Here, we instead use a simpler mapping
based on the one-to-one correspondence between the boson and fermion 
configurations, as described earlier by 
Toreblad {\it et al.}~\cite{toreblad2004}. The bose ``condensate'', 
$\vert N0000\cdots\rangle$,  corresponds to 
the MDD in the fermion case. Other configurations can be 
obtained as single-particle excitations out of the condensate.
Table~\ref{confs} shows a few examples of these configurations, here 
for the simple case of six particles. 
\begin{table}[h]
\caption{Examples of corresponding configurations for six fermions and
bosons. $M$ is the angular momentum of the configuration. 
For six fermions the angular momentum of the MDD is 15.}
\begin{tabular}{llcl}
\hline
$M_{\rm fermion}$ &fermion state& $M_{\rm boson}$& boson state\\
\hline
15+0 & $\vert 1111110000\rangle$ & 0 & $\vert 600000\rangle$ \\
15+4 & $\vert 1101111000\rangle$ & 4 & $\vert 240000\rangle$ \\
15+4 & $\vert 1111001100\rangle$ & 4 & $\vert 402000\rangle$ \\
15+15 & $\vert 1000111110\rangle$ & 0 & $\vert 100500\rangle$ \\
\hline
\end{tabular}
\label{confs}
\end{table}
Figure~\ref{amps} compares the amplitudes of the most important
configurations for the two-vortex state of 20 bosons with
$M=34$ to those of 20 fermions with $M=224$. 
For bosons, the state is at the yrast line, while for fermions it is the second
excited state. The yellow columns show the cumulative overlap between the 
boson and fermion states. The configurations are shown at the right, 
using the notation for bosons.
The figure shows that for both kinds of particles,
the same configurations are important. The actual amplitudes differ,
but the qualitative similarity of the states is clearly seen,
especially regarding  the signs of the different terms contributing to the
linear combination of Slater determinants in the many-body state. 

\subsection{Density and vorticity}
\label{sec_densvor}

Any exact solution of interacting particles
in circularly symmetric potential must have 
a density with circular symmetry. Thus, only the radial
density distribution is relevant. In the LLL
the density can be simply determined from the 
occupancies of the single-particle levels
\be
n(r)=\sum_\alpha \vert C_\alpha\vert^2 \sum_m n_{\alpha m}
\vert\psi_m(r,\theta)\vert^2,
\ee
where $C_\alpha$ is the amplitude of the configuration $\alpha$,
$n_{\alpha m}$ the occupation number of the single
particle state $m$ in configuration $\alpha$, and
$\psi_m$ the single-particle state of Eq.~(\ref{spstate}).
We note that the same information is contained in the occupancy of the 
single-particle angular momentum states:
\be
D(m)=\sum_\alpha \vert C_\alpha\vert^2 n_{\alpha m}.
\label{dm}
\ee 
The restriction to the LLL makes it simple to determine
the current density
\be
j_\theta(r)=\sum_\alpha \vert C_\alpha\vert^2 \sum_m n_{\alpha m}
\frac{m}{r}\vert\psi_m(r,\theta)\vert^2,
\label{current}
\ee
and the velocity field $j_\theta(r)/n(r)$. Naturally, the current
density as well as the velocity field have the  circular symmetry of the
underlying Hamiltonian. 

We can also use the vorticity of the velocity field
\be
\nabla\times \left(\frac{j_\theta(r)}{n(r)}\right){\bf e}_\theta
=\frac{\partial}{\partial r} 
\left(\frac{j_\theta(r)}{n(r)}\right){\bf e}_z
\label{vorticity}
\ee
to give information about the vortex formation.
If the vortex is at the center of the trap, naturally the density of the
single-vortex state is rotationally invariant. 
However, if one or more vortices are off-center, their 
characterization is not as straight-forward.

\begin{figure}[h]
\includegraphics[width=0.95\columnwidth]{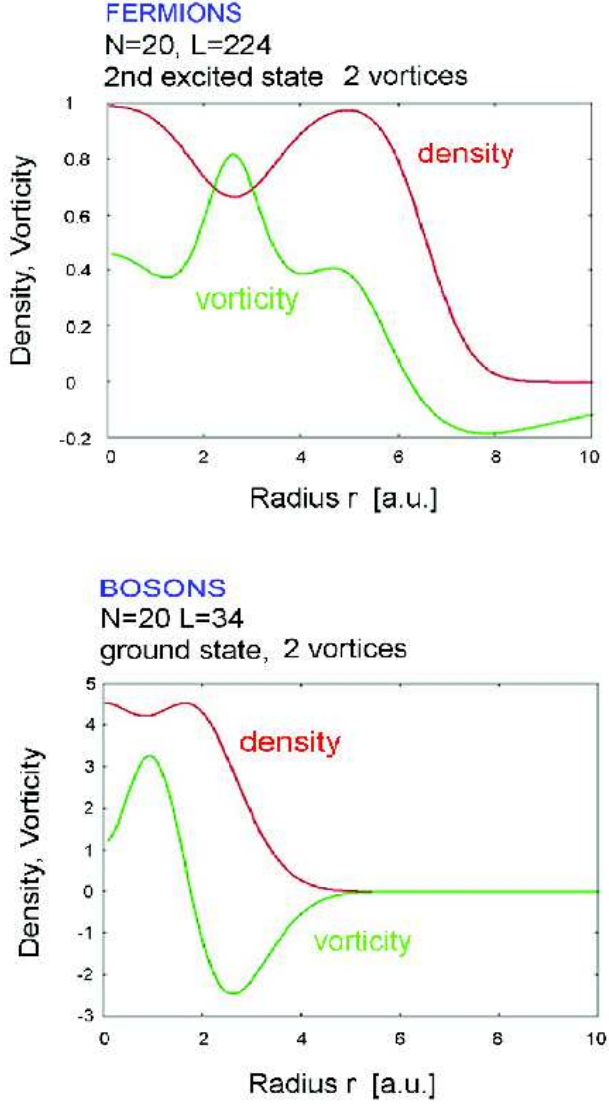}
\caption{Density ({\it red lines}) and vorticity ({\it green lines})
in a fermionic ({\it upper panel}) 
and bosonic ({\it lower panel}) system, respectively. 
For $N=20$ particles, 
the fermionic two-vortex state at $M=190+34$ 
is compared to its bosonic counterpart at $M=34$.
}
\label{pic_vorticity}
\end{figure}

For example, the two-vortex solution, as displayed in 
Fig.~\ref{pic_vorticity}, shows a clear 
maximum in vorticity where the radial density profile has a local minimum. 
The profiles of the density and vorticity for bosons and fermions, 
respectively, appear rather similar, with the main difference that 
the fermionic cloud extends to larger radius as a 
simple consequence of the Pauli principle. 

\section{Trial wave functions for vortex states and fermion-boson relations}
\label{sec_trial}

The fermion wave function for the maximum density droplet (MDD)
is a single determinant, see Eq.~(\ref{MDD}), 
where the number of singly occupied states equals the number of particles. 
Using the single-particle states of the LLL, Eq.~(\ref{spstate}),
now in terms of the complex coordinates $z_j=x_j+iy_j$, 
this takes the form of the Laughlin wave function~\cite{laughlin1983}
which, apart from the normalization, is written as
\begin{equation}
\Psi_{q}^F=\prod_{i<j}^N (z_i-z_j)^q\exp\left(-\sum_k\vert
  z_k\vert^2/4\right),
\label{laughlin}
\end{equation}
where $q=1$ for MDD and in general an odd integer for fermions.
For the MDD, we use the notation $\Psi _{1}^F=\Psi ^F_{MDD}$.
When the angular momentum is increased, there is no simple 
analytic formula for the exact state. However, it has been shown 
that at $3M_{MDD}$, the
next Laughlin state with $q=3$ corresponding to the filling factor 
$\nu=1/3$ for the fractional quantum Hall effect is a good
{\it approximation} to the exact wave function~\cite{manninen2001}.
For angular momenta in between the Laughlin states, no simple
analytic expressions for the wave function exist.

Rather than providing accurate analytic estimates for the many-body 
wave function, our aim here only is to reach
a physical understanding of vortex formation and crystallization at 
high angular momenta. To this end, let us first consider the simplest 
case of a single vortex.
Bertsch and Papenbrock~\cite{bertsch1999} have shown that for weakly
interacting bosons, a single vortex at the center of the condensate
can be described as
\be
\Psi_{1v}^B=\prod_i^N (z_i-z_0)\exp\left(-\sum_k\vert
  z_k\vert^2/4\right),
\label{bp}
\ee
where $z_0=(z_1+z_2+\cdots z_N)/N$ is the center-of-mass coordinate.
We can use the same Ansatz for fermions and write~\cite{manninen2001}
\be
\Psi_{1v}^F=\prod_i^N (z_i-z_0)\Psi_{MDD}^F.
\label{eq1v}
\ee
Note that the only difference between the boson and fermion states
is the additional product $\prod_{i<j}^N (z_i-z_j)$ which makes the 
fermion state antisymmetric. We have tested this approximation in
the case of six electrons. In this case, the overlap between the 
trial wave function Eq.~(\ref{eq1v}) and the exact result (restricted to LLL)
for Coulomb interaction is 98.5 \%. 
Even when the restriction to LLL is abandoned and 
higher Landau levels are included,  the overlap was found~\cite{manninen2001} 
to be 90\%~.

The most important configuration of the Bertsch-Papenbrock state 
Eq.~(\ref{bp}) for, say six, fermions is $\vert 01111110000\rangle$, 
i.e. one electron is
removed from the angular momentum $m=0$ state and lifted to the first 
empty angular momentum state. We notice that this is the smallest
angular momentum state that can have an empty state at $m=0$. For six
electrons, $M_{MDD}+N=15+6=21$. 
For smaller values of $M$ the empty state is at a higher single-particle
state. For example, for $M=M_{MDD}+1$ the only possible configuration 
is $\vert 11111010000\rangle$ and the corresponding wave function is
$x_0\Psi_{MDD}^F$. When the angular momentum is increased, the hole
in the Fermi sea moves to lower angular momentum until it reaches the 
origin. We will call these states as one-vortex states. When the 
hole is not at $m=0$ the vortex is not localized at the origin, 
but de-localized in a ring at a radius which depends
on the angular momentum of the hole.

In the case of many vortices, simple analytic
approximations  for the states are not known. 
However, we can guess the most 
important configuration with the help of the following
arguments~\cite{toreblad2004}.
First we notice that for a large number of particles we can replace
the center of mass in Eq.~(\ref{bp}) with the origin, i.e. with a
fixed point. In the same spirit we can assume several localized
vortices at fixed points. When the number of vortices is small, their
geometrical arrangement will be on a ring. In this case 
the wave function corresponding to Eq.~(\ref{bp}) would be
\begin{eqnarray}
\Psi_{kV} &=&\prod_{j_1}^N (z_{j_1}-ae^{i\alpha_1})\times \cdots\times
\prod_{j_k}^N (z_{j_k}-ae^{i\alpha_k}) \Psi_{MDD} \nonumber\\
&=& \prod_j^N (z_j^k-a^n)\Psi_{MDD}~,
\end{eqnarray}
where $k$ is the number of vortices, $a$ is distance of the
vortices from the origin and $\alpha_j=2\pi j/k$. 
Clearly, the above wave function does not have a good angular momentum.
Projecting to a good angular momentum means collecting out states
with a given power of $a$. We get a state
\be
\Psi_{kV}=a^{k(N-K)}{\cal{S}}\left(\prod_j^K z_j^k\right)\Psi_{MDD}
\label{vgen}
\ee
which now
corresponds to a good angular momentum $M=M_{MDD}+kK$.
Expanding the polynomial in Eq.~(\ref{vgen}) shows that the most
important configuration has the form
\begin{equation}
\vert \underbrace{1111}_{N-K}000\underbrace{1111111111}_{K}0000000\rangle ,
\end{equation}
where the number of 
adjacent zeros between the one's equals the number of vortices,
and the number of ones after the zeros equals to $K$. 

In the case of bosons, a trial wave function can be constructed in a
similar fashion, the only difference being that the MDD wave function,
Eq.~(\ref{laughlin}), is replaced by the Bose condensate,
$\exp(-\sum\vert z_k\vert^2)$. This means that the trial wave
functions for bosons and fermions are similar apart from the
product in Eq.~(\ref{laughlin}), which makes the fermionic state 
antisymmetric. 

Though originally developed to describe fractional quantum Hall
states, a tool that has shown useful to describe the high angular momentum 
states of rotating bose condensates~\cite{cooper1999,wilkin2000,viefers2000,chang2005,regnault2003,regnault2004}
is the composite fermion model of Jain~\cite{jain1998}. 
In this method, the wave function for any $M$
is determined first using also higher Landau levels and then
projected to the lowest Landau level. 
While it would be expected to be applicable in the regime 
$M\sim N ^2$,  it was more recently extended successfully to 
describe angular momenta even before the unit 
vortex~\cite{viefers2000,korslund2006}.

The construction of the composite
fermion state is complicated due to the projection to the LLL
and does not easily reveal the nature of the solution. 
Nevertheless, the composite fermion 
picture shows that there is a relation 
between the states of the spinless fermions and bosons.
In fact, it has been shown~\cite{ruuska2005}
that in the analytically solvable model of harmonic interparticle 
interactions, this relation between the bosonic and
fermionic solutions is exact and the states can be written
as a product of a symmetric homomorphic polynomial
$P_M$ and the boson or fermion condensate
\smallskip
\noindent 
for fermions,
\be 
\Psi_{M_{MDD}+M}=P_M \prod(z_i-z_j)\exp\left(-\sum_k\vert
  z_k\vert^2/4\right) 
\ee
and for bosons,
\be
\Psi_M=P_M \exp\left(-\sum_k\vert
  z_k\vert^2/4\right)~. 
\ee
The only difference between the boson and fermion
states is then the product that makes the fermion state
antisymmetric. (In the case of the Laughlin
state with $q\ge 3$ the symmetric polynomial is simply
$P_{(q-1)N(N-1)/2}=\prod (z_i-z_j)^{q-1}$).

\section{conclusions}
\label{sec_conclude}

At extreme angular momenta, a few particles in a harmonic trap 
crystallize, indpendent of their fermionic or bosonic nature. 
In much analogy to what was observed in quantum dots at strong magnetic
fields~\cite{maksym1990,maksym1996}, this 
crystallization is apparent from the 
regular oscillations in the quantum many-body spectrum: 
Cusps at the yrast line, and their periodicity in angular momentum, 
are easily understood in terms of the simple geometries of these 
so-called Wigner molecules, being of either fermionic or bosonic 
nature~\cite{manninen2001,romanovsky2004}.

At moderate angular momenta, however, it is 
known that vortices may form, independent of the system being bosonic or
fermionic.  In a harmonic confinement, these ``holes'' that are 
penetrating the rotating quantum system, arrange in simple polygonal 
structures, very similar to those of the localized Wigner crystallites 
of particles. They lead to similar cusp states in the yrast line, 
with oscillation periods simply following the underlying symmetry 
of the ``vortex crystal''. 
We showed how this mapping could be understood in terms of  
particle-hole duality, holding for the bosonic as well as the fermionic 
case~\cite{manninen2005}.

\section*{Acknowledgements}
We thank D. Pfannkuche, B. Mottelson, S. Viefers, J. Jain,  
H. Saarikoski and E. R\"as\"anen for rewarding discussions. 
Financial support from the Swedish Foundation for Strategic Research
and the Swedish Research Council, from NordForsk as well as the 
Academy of Finland is gratefully acknowledged.

\end{document}